\documentclass{elsart5}

 \usepackage{graphicx}
\usepackage{amssymb}

\begin{document}

\begin{frontmatter}

%\title{Title\thanksref{tit1}}
%\thanks[tit1]{Title footnote}
\title{Internal magnetic field effect on magnetoelectricity \\in orthorhombic 
$R$MnO$_{3}$ crystals}

\author[]{K. Noda\corauthref{cor1}} %\thanksref{label1}}
\ead{n-kohei@sophia.ac.jp}
%\thanks[label1]{author footnote}
\corauth[cor1]{Tel.:+81-3-3238-3399;Fax:+81-3-3238-3430}
\author[]{M. Akaki}
\author[]{F. Nakamura}
\author[]{D. Akahoshi}
\author[]{H. Kuwahara}
%\address{The Department of Physics, Sophia University, 7-1 Kioi-cho, Chiyoda-ku, Tokyo, 102-8554, Japan.}
\address{Department of Physics, Sophia University, 7-1 Kioi-cho, Chiyoda-ku, Tokyo 102-8554, Japan.}
%\address[aff2]{Address}
%\thanks[label2]{aff footnote}
%\received{12 June 2005}
%\revised{13 June 2005}
%\accepted{14 June 2005}
%use optional labels to link authors explicitly to addresses:

%\author{}
%\address{}

\begin{abstract}
We have investigated the role of the 4$f$ moment on the magnetoelectric (ME) effect of orthorhombic $R$MnO$_{3}$ ($R$=rare earth ions). 
In order to clarify the role of the 4$f$ moment, we prepared three samples: (Eu,Y)MnO$_{3}$ without the 4$f$ moment, TbMnO$_{3}$ with the anisotropic 4$f$ moment, and (Gd,Y)MnO$_{3}$ with the isotropic 4$f$ moment. 
%The average ionic radius of $R$ site in these samples is same as that of TbMnO$_{3}$. 
The ferroelectric behaviors of these samples are different from each other in a zero magnetic field. (Eu,Y)MnO$_{3}$ and (Gd,Y)MnO$_{3}$ show the ferroelectric polarization along the $a$ axis in the ground state, while TbMnO$_{3}$ shows it along the $c$ axis. 
Such difference may arise from the influence of the anisotropic Tb$^{3+}$ 4$f$ moment.
% which works as an internal magnetic field along the $a$ axis. 
The direction of the ferroelectric polarization of $R$MnO$_{3}$ is determined by the internal magnetic field arising from the 4$f$ moment.
\end{abstract}

%%%%%%%%%use  the \KEY command at the begin of keyword text%%%%%%%%%
\begin{keyword}
\PACS 75.80.+q \sep 71.45.Gm \sep 77.84.Bw \sep 77.80.Fm \sep 75.30.-m \sep 75.50.Ee \sep
\KEY Multiferroics \sep Magnetoelectric effect \sep Ferroelectrics \sep Antiferromagnetism
\end{keyword}
%Please supply one or more relevant PACS-1996 classification codes 
%(http://publish.aps.org/PACS/96pacs.htmland) and about 5 keywords 
%of your own choice for indexing purposes. 
%You can see a list of already used keywords for JMMM at 
%http://authors.elsevier.com/JournalDetail.html?PubID=505704&Precis=KIND

\end{frontmatter}

%\section{Introduction}\label{}
A strong correlation between dielectric and magnetic properties, so called magnetoelectric effect (ME), has attracted a revived interest. 
The series of orthorhombic $R$MnO$_{3}$ ($R$=rare earth ions) containing TbMnO$_{3}$ famous as the ``magnetic-field-induced electric polarization flop"\cite{kimura} is a parent material of the colossal magnetoresistance (CMR) manganite. 
Recently, the noncollinear transverse spiral antiferromagnetic (AF) order of Mn 3$d$ spins is observed between the $A$-type (EuMnO$_{3}$) and the $E$-type (HoMnO$_{3}$) AF order.\cite{Kimura-2,Kenzelmann,Arima} 
Such unconventional magnetic order is caused by the competition between the ferromagnetic nearest neighbor (NN) interaction and the AF next NN interaction due to the relatively large orthorhombic distortion. 
Such noncollinear transverse spiral magnetic structure causes the ferroelectric polarization of $R$MnO$_{3}$.\cite{Kenzelmann,Arima,Katsura,Mostovoy} 

In previous work, we have investigated the ME effect in (Eu$_{0.595}$Y$_{0.405}$)MnO$_{3}$ ((Eu,Y)MnO$_{3}$) without 4$f$ moment.\cite{Noda} 
Even if the average ionic radius of $R$ site in (Eu,Y)MnO$_{3}$ is the same as that of TbMnO$_{3}$, the direction of the ferroelectric polarization of the ground state of (Eu,Y)MnO$_{3}$ is different from that of TbMnO$_{3}$ in a zero magnetic field: (Eu,Y)MnO$_{3}$ shows the ferroelectric polarization along the $a$ axis ($P_{a}$), while TbMnO$_{3}$ shows $P_{c}$. 
It is considered that this difference is attributed to the magnetic property of each $R$ ion.  

In this paper, we have revealed the role of the rare-earth 4$f$ moment on the ME effect of orthorhombic $R$MnO$_{3}$. In order to investigate the role of the 4$f$ moment, we prepared three samples: (Eu,Y)MnO$_{3}$ without the 4$f$ moment, TbMnO$_{3}$ with the anisotropic 4$f$ moment, and (Gd$_{0.69}$Y$_{0.31}$)MnO$_{3}$ ((Gd,Y)MnO$_{3}$) with the isotropic 4$f$ moment. 
The average ionic radius of $R$ site in these samples is fixed to be same as that of TbMnO$_{3}$. 
%The dielectric and magnetic properties of these samples suggest that the rare-earth 4$f$ moment works as the internal magnetic field and determines the direction of the ferroelectric polarization in a zero magnetic field. 

%%%%%%%%%%%%%%%%%%%% Kondo in title, abstract and/or keywords %%%%%%%%%%%%%%%%

  \begin{figure*}[ht]
  \begin{center}
\includegraphics[scale =0.58]{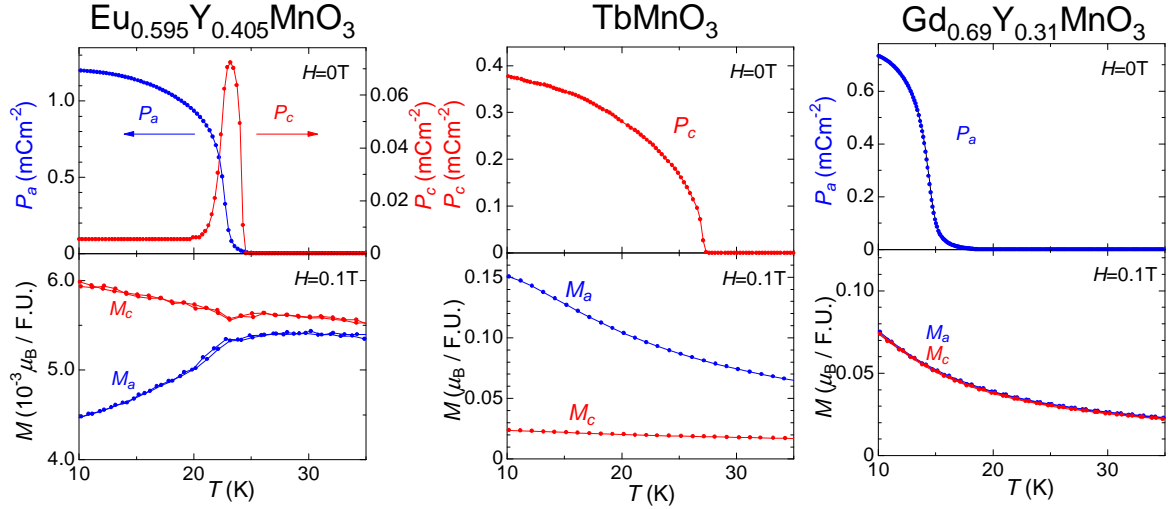}
\caption{Temperature dependence of the ferroelectric polarization (upper) and magnetization (lower) of the (Eu,Y)MnO$_{3}$ (left panel), TbMnO$_{3}$ (center panel), and (Gd,Y)MnO$_{3}$ (right panel). 
}
  	\label{fig-1}
  \end{center}	
  \end{figure*}
    
%%%%%%%%%%%%%%%%%%%%%%%%%%%%%%%%%%%%%%%%%%%%%%%%%%%%%%%%%%%%%%%%%%%%%%%%%%%%%%

%\section{Experiment}
%We prepared (Eu,Y)MnO$_{3}$, TbMnO$_{3}$ and (Gd,Y)MnO$_{3}$.
%The average ionic radius of the $R$ site of these samples is fixed to be same as that of TbMnO$_{3}$. 
The compositional ratio of the Eu$^{3+}$ to Y$^{3+}$ ions and that of Gd$^{3+}$ to Y$^{3+}$ were obtained by a calculation based on the Shannon's ionic radius table.\cite{shanon} 
The single crystal sample was grown by the floating zone method. 
We performed x-ray-diffraction and rocking curve measurements on the resulting crystal at room temperature, and confirmed that the samples have the orthorhombic $Pbnm$ structure without any impurity phases or any phase segregation. 
All specimens used in this study were cut along the crystallographic principal axes into a rectangular shape by means of x-ray back-reflection Laue technique. 
The measurement of the temperature dependence of the spontaneous ferroelectric polarization was carried out in a temperature-controllable cryostat.
%The dielectric constant was determined with an LCR meter (Agilent, 4284A). 
%After the sample had been cooled under a poling electric field of 300$\sim$500 kV/m, 
The spontaneous electric polarization was obtained by the accumulation of a pyroelectric current. 
%while it was heated at a rate of 4K/min. 
The magnetization was measured using a commercial apparatus (Quantum Design, PPMS).

%\section{Result and Discussion}

At first, we focus on (Eu,Y)MnO$_{3}$without 4$f$ moment. 
In a zero magnetic field, this compound shows two distinct ferroelectric phases with $P_{a}$ ($T$$\leq$23K) and $P_{c}$ (23K$\leq$$T$$\leq$25K) (See Fig.\ref{fig-1} left panel). 
%The ground state of (Eu,Y)MnO$_{3}$ has $P_{a}$. 
The magnitude of the magnetization is about 10$^{-3}$$\sim$10$^{-4}$ $\mu$$_{\rm B}$$/$F.U. in $H$=0.5T. 
The behavior of magnetization is directly reflecting the behavior of the Mn 3$d$ spins, because this compound is free from 4$f$ moment. 
In previous work, we have demonstrated the ferroelectric polarization switching between $P_{a}$ and $P_{c}$ by application of the magnetic field: For example, the $P_{a}$ changes to the $P_{c}$ by the application of the magnetic field parallel to the $a$ axis ($H_{a}^{\rm ext}$).\cite{Noda} 
This switching behavior is caused by the flop of the transverse spiral AF plane of Mn 3$d$ spins, which can be explained in terms of the inverse Dzyaloshinskii-Moriya interaction.\cite{Mostovoy} 

In the case of TbMnO$_{3}$, the dielectric and magnetic properties are different from those of (Eu,Y)MnO$_{3}$, although the average ionic radius of $R$ site is the same (See Fig.\ref{fig-1} center panel). 
TbMnO$_{3}$ shows only $P_{c}$ in a zero magnetic field. 
The magnitude of the magnetization is about 10$^{1}$$\sim$10$^{2}$ times as large as that of 4$f$-moment free (Eu,Y)MnO$_{3}$, because of the influence of the Tb$^{3+}$ 4$f$ moment. 
Introducing the 4$f$ moment enhances the magnitude of the magnetization along the $a$ axis ($M_{a}$) compared with the other axes. 
This result suggests that the 4$f$ moment of TbMnO$_{3}$ is anisotropic, and the internal magnetic field due to the 4$f$ moment seems to be parallel to the $a$ axis ($H_{a}^{\rm int}$). 
In the case of 4$f$-moment free (Eu,Y)MnO$_{3}$, the ground state has $P_{a}$, and the direction of the ferroelectric polarization is changed from the $a$ axis to the $c$ axis by the application of $H_{a}^{\rm ext}$, as mentioned above. 
Therefore, in TbMnO$_{3}$, $P_{c}$ is stabilized by the application of $H_{a}^{\rm int}$, like the case of 4$f$-moment free (Eu,Y)MnO$_{3}$ in $H_{a}^{\rm ext}$. 
%if $H_{a}^{\rm int}$ acts as $H_{a}^{\rm ext}$, $H_{a}^{\rm int}$ influences the spiral AF structure of Mn 3$d$ spins, and 
Hence, the difference of the direction of the ferroelectric polarization between 4$f$-moment free (Eu,Y)MnO$_{3}$ and anisotropic 4$f$-moment TbMnO$_{3}$ can be understood from the scenario that $H_{a}^{\rm int}$ acts as $H_{a}^{\rm ext}$. 
%Such $H_{a}^{\rm int}$ stabilizes the $P_{c}$, which is consistent with the case of the $P_{c}$ in $H_{a}^{\rm ext}$ for the 4$f$-moment-free compound, (Eu,Y)MnO$_{3}$.

In the case of (Gd,Y)MnO$_{3}$, the $P_{a}$ only appears in a zero magnetic field (See Fig.\ref{fig-1} right panel).
The magnitude of the magnetization is nearly the same as that of TbMnO$_{3}$. 
However, in this compound, no significant difference is observed between the temperature dependence of $M_{a}$ and $M_{c}$, in contrast to the case of TbMnO$_{3}$. 
Therefore, the 4$f$ moment of this compound is isotropic.
%, which does not cause the anisotropic internal magnetic field. 
As a result, the effective internal magnetic field does not exist. 
% does not feel  induced by the 4$f$ moment. 
Consequently, the direction of the ferroelectric polarization of this compound is the same as that of 4$f$-moment free (Eu,Y)MnO$_{3}$. 

From these results, we conclude that, in $R$MnO$_{3}$ crystals, the Mn 3$d$ spins are indispensable for realizing the ferroelectric polarization, while the 4$f$ moment is not. The magnetic easy axis of the 4$f$ moment determines the direction of the ferroelectric polarization through the channel of the internal magnetic field, even in a zero external magnetic field. 
The results obtained in this experiment should provide an improved understanding of the mechanism of the magnetoelectric effect in $R$MnO$_{3}$ crystals.@

%\section{Summary}

%\appendix 
%\section{}

\end{document}